\documentclass{pasj00}
\Received{$\langle$reception date$\rangle$}
\Accepted{$\langle$acception date$\rangle$}
\Published{$\langle$publication date$\rangle$}
\SetRunningHead{Honma et al.}{Super-resolution imaging for radio
interferometer}

\usepackage{times}

\begin{document}

\title{Super-resolution imaging with radio interferometer using sparse
modeling}
\author{
Mareki \textsc{Honma},\altaffilmark{1,2}
Kazunori \textsc{Akiyama},\altaffilmark{1,3}
Makoto \textsc{Uemura},\altaffilmark{4}
\& 
Shiro \textsc{Ikeda},\altaffilmark{5,6}
}
\altaffiltext{1}{Mizusawa VLBI Observatory, NAOJ, Mitaka, 181-8588}
\altaffiltext{2}{Department of Astronomical Science, Graduate University for Advanced Studies (SOKENDAI), Mitaka, 181-8588}
\altaffiltext{3}{Department of Astronomy, The University of Tokyo, Hongo, Bunkyo, Tokyo 113-0033}
\altaffiltext{4}{Astrophysical Science Center, Hiroshima University,
Higashi-Hiroshima, 739-8526}
\altaffiltext{5}{Institute of Statistical Mathmatics, Tachikawa, 190-8562}
\altaffiltext{6}{Department of Statistical Science, Graduate University
for Advanced Studies (SOKENDAI), Tachikawa 190-8562}
\email{mareki.honma@nao.ac.jp}

\maketitle

\begin{abstract}
We propose a new technique to obtain super-resolution images with radio
 interferometer using sparse modeling.
In standard radio interferometry, sampling of ($u$, $v$) is quite often
 incomplete and thus obtaining an image from observed visibilities
 becomes an underdetermined problem, and a technique so-called
 {``}zero-padding{''} is often used to fill up unsampled grids in ($u$,
 $v$) plane, resulting in image degradation by finite beam size as well
 as numerous side-lobes.
In this paper we show that directly solving such an underdetermined
 problem based on sparse modeling (in this paper LASSO) avoids the
 above problems introduced by zero-padding, leading to
 super-resolution  images in which structure finer than the standard
 beam size (diffraction limit) can be reproduced.
We present results of one-dimensional and two-dimensional simulations of
 interferometric imaging, and discuss its implications to super-resolution
 imaging, particularly focusing on imaging
 of black hole shadows with millimeter VLBI.
\end{abstract}

\section{Introduction}

\subsection{toward higher resolution}

Pursuit of higher angular resolution is fundamental to the development
of modern observational astronomy because higher angular resolution provides
better knowledge of detailed structure of astronomical objects.
At any wavelength of electromagnetic wave, angular resolution
(or often referred to as {``}beam size{''} in radio astronomy, and
{``}diffraction limit{''} in optical astronomy) is simply given by
\begin{equation}
\label{eq:beamsize}
 \theta \approx \lambda/D,
\end{equation}
where $\theta$ is the angular resolution, $\lambda$ is the observing wavelength, and $D$ is the diameter of the telescope.
For a given wavelength of $\lambda$, it is essential to have
a larger telescope to obtain higher angular resolution.
However, building a large telescope is highly demanding in terms of
technology as well as cost, and practically there is a severe limit of
the size of a single-dish telescope.

Interferometer is an alternative approach to obtain high angular
 resolution by synthesizing a large telescope that cannot be realized as
 a stand-alone single telescope.
Through the whole electromagnetic spectral window, radio interferometer
is most successful for synthesizing a huge telescope and consequently a
 sharp beam (e.g., \citet{TMS01}).
In particular,
VLBI (Very Long Baseline Interferometer),
which utilizes intercontinental baselines or sometimes even space baselines,
provides the highest angular resolution ever obtained in the history of
 astronomy.
VLBI with Earth-size baselines at cm wavelength readily provides an
angular resolution as small as 0.1 mas.
However, astronomical objects still have structures smaller than 
the angular resolution available with VLBI, and there is definitely need
 of higher angular resolution.
Unfortunately, the maximum baseline of VLBI array is basically limited by
the Earth size, except in case of space VLBI projects such as 
VSOP/HALCA \citep{Hirabayashi98} and RadioAstron \citep{Kardashev13}.
Once the array size is reached at the maximum, the only way
left for improving resolution is to go to shorter wavelength.

In order to achieve higher angular resolution, a lot of efforts
have been made to build an international mm/sub-mm VLBI array (e.g.,
EHT: Event Horizon Telescope, see e.g., Doeleman et al. 2009).
One of the major goals of high-angular resolution observations with EHT
is to resolve event horizon scale of nearby super-massive
black holes such as Sgr A* and M87.
For doing this, a resolution of 10 $\mu$as scale is required.
However, going to a short wavelength (or a high frequency) is
a big challenge, and the angular resolution cannot be improved infinitely 
by shortening the wavelength.
For instance, practical limit for sub-mm VLBI is thought to be $\sim $0.8 mm
(or 350 GHz), considering available sites and stations on the Earth.
VLBI with a 0.8 mm wavelength and a 9000 km baseline would provide an
angular resolution of $\sim$ 20 $\mu$as.
However, this resolution may not be enough to resolve the black hole
structure in the vicinity of event horizon depending on black hole
size, and there would be a further need of improvement of angular
resolution.

Super-resolution, which provides a higher resolution than the
standard diffraction limit of $\lambda/D$, is an alternative and
attractive approach to break the limit of angular resolution imposed by
the Earth size and the maximum observing frequency.
Super-resolution is equivalent to expanding array size
and/or going to higher frequency, and thus it will provide a
significant impact on high-resolution astronomy such as black hole
imaging and other fields in astronomy as well.

\subsection{Basics of radio interferometer and sparse modeling}

VLBI, for which we would like to discuss the possibility of
super-resolution imaging, is a type of radio interferometer that is
in principle the same to connected arrays.
The raw output of a radio interferometer is cross-power spectrum, 
which is also referred to as visibility $\mathcal{V}(u,v)$ as a function
 of spatial frequency ($u$, $v$).
The spatial frequency corresponds to projected baseline normalized with
the observing wavelength (i.e., $u\equiv U/\lambda$ and $v\equiv
V/\lambda$, where  $U$ and $V$ are physical lengths of the projected
baseline).
As is widely known, interferometric images of astronomical sources
$I(x,y)$ can be obtained by
 two-dimensional Fourier transform of observed visibilities $\mathcal{V}(u,v)$ as,
\begin{equation}
\label{eq:2D-FT}
 I(x, y) = \int\!\!\! \int \mathcal{V}(u,v)\, e^{-2\pi i (ux +vy)}\, du dv.
\end{equation}
By definition of Fourier transform, the integration for $u$ and $v$
should be conducted over infinite range (i.e., from $-\infty$ to
$\infty$).
However, practically this is impossible: first, the array size is finite
and there should be a limit of baseline length, beyond which no visibility
can be sampled.
Secondly, there must be space between antennas, which causes
unsampled {`}holes{'} in the $(u, v)$ plane.
Such an incomplete ($u$, $v$) coverage always causes
{``}underdetermined problem{''} in the radio interferometer equation
such as equation (\ref{eq:2D-FT}).
Usually this problem is solved by filling unsampled visibilities with
zero, which is referred to as {``}zero-padding{''} in the present paper.
It is this process
that causes a finite beam size in the resultant images and hence limits
the image resolution.
We note that in classical Fourier transformation, all unmeasured visibilities are implicitly set to zero, and we refer to this loosely as the {``}zero-padding{''} problem, while normally the term is used to describe the insertion of a block of zeroes to match the grid size to 2$^N$ and/or to interpolate the spectrum.

On the other hand, there are different approaches to reconstruct radio 
interferometer images
such as Maximum Entropy Method (MEM) and/or model fitting, which are
different from the standard imaging synthesis described above in the
sense that these methods do not utilize Fourier transformation with
zero-padding but instead select the model which describes observed
visibilities best using other criteria (as the problem is
underdetermined and has infinite number of solutions) .
As seen later in this paper, such an approach to
select the best solution can provide a resolution higher
than the standard imaging provided an appropriate selection is done.
In fact, there have been several discussions on possibility of
super-resolution by MEM and model fitting to visibilities (e.g.,
Cornwell \& Evans 1985).

Meanwhile, sparse modeling is recognized as a new and powerful approach to
solve an underdetermined problem by introducing sparsity of the
solution.
Recent developments of compressive sensing (Donoho 2006, Candes \&
Tao 2006)
and LASSO (Least Absolute Shrinkage and Selection Operator,
Tibshirani 1996)
provide mathematical tools of sparse modeling, which may be applied to
radio interferometry imaging.
In fact, pioneering works such as \citet{Wiaux09} and \citet{Li11}
discussed application of compressive sensing and demonstrated its power
in radio interferometer imaging based on comparisons with standard
CLEAN images.
In this paper, in order to evaluate the impact of sparse modeling technique
on super-resolution imaging, we apply LASSO to simulated data
and discuss its application to imaging of black hole shadows.

\section{One-dimensional consideration on super-resolution}

\subsection{Standard imaging}

\begin{figure*}[t]
\begin{center}
       \FigureFile(130mm,130mm){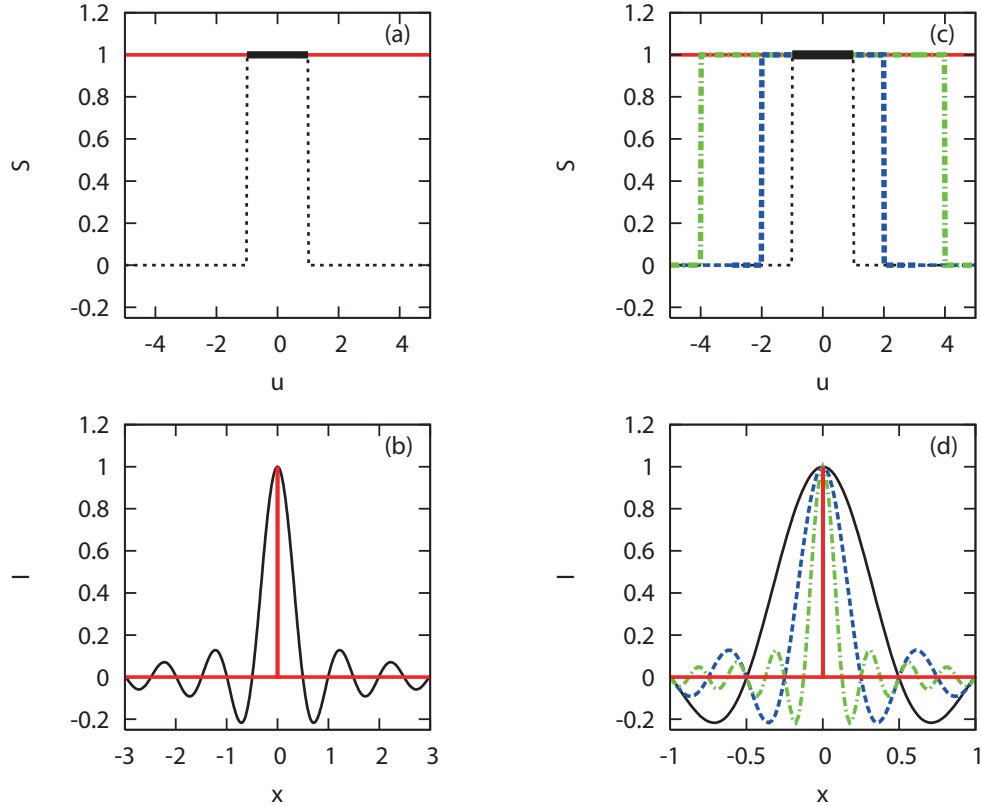}
\end{center}
\caption{One-dimensional examples of visibilities and images for a point source
observed with finite spatial frequency coverage.
Top-left panels show the visibility for a point source, and bottom-left 
panels are its Fourier transform (image). Note that the peak intensity
 in the bottom panels are scaled to unity.
Red line in (a) corresponds to
 a full sampling of spatial frequency for $|u|<\infty$, and red
 line in (b) is its Fourier transform.
 Thick black line and curve in (a) and (b) corresponds to a limited
 sampling of spatial frequency for $|u|<1$, showing that such a
 limited spatial frequency causes finite beam size as well as numerous
 side-lobes.
Top-right (c) and bottom-right (d) are similar pairs of visibilities
 and images, with different spatial frequency coverage: 
 black curves for $|u|<1$, blue for $|u|<2$, green for $|u|<4$, and
 red for $|u|<\infty$.
Note that all the four cases satisfy $\mathcal{V}=1$ for $|u|<1$.
}
\end{figure*}

Here we start our discussion on super-resolution by reviewing how the
synthesized beam is generated by limited $(u ,v)$ coverage.
To clearly illustrate the effect of finite $(u ,v)$ coverage
on the beam size (and how this effect can be avoided to obtain
super-resolution image), here we consider a simple one-dimensional
problem, where visibility $\mathcal{V}(u)$ and intensity $I(x)$ is related
with each other through one-dimensional Fourier transform:
\begin{equation}
\label{eq:1D-FT}
  I(x) = \int \mathcal{V}(u)\, e^{-2\pi i ux}\, du.
\end{equation}
Suppose that there is a point source at $x=0$, which is described
as $I(x)=\delta(x)$ ($\delta(x)$ is Dirac's delta function),
and also suppose that visibility $\mathcal{V}(u)$ is uniformly sampled
for spatial frequency $u$ between $-u_{\rm max}$ and $u_{\rm max}$.
Figures 1 (a) and (b) show the case for $u_{\rm max}=1$: in figure 1(a),
the true visibility for the point source is $\mathcal{V}(u)=1$ for any $u$, but
actual sampling is done only between $u=-1$ and $u=1$.
For spatial frequency at $|u|\ge 1$, there is no observation and
hence no information is available on the source visibility.
On the other hand, figure 1(b) shows the Fourier transform of the true
visibility ($\mathcal{V}(u)=1$ for any $u$) and the observed visibility
($\mathcal{V}(u)=1$ only for $|u|\le 1$ and 0 for other range of $u$).
As is widely known, the latter case generates an extended beam in 
the form of ${\rm sinc}(x)(=\sin(x)/x)$ function, with infinite number
of side-lobes as seen in figure 1(b).
The size of the main beam is $\sim 1/u_{\rm max}=\lambda/B_{\rm max}$,
which is consistent with equation (\ref{eq:beamsize}).
This is the standard approach to obtain an image in radio interferometry
synthesis, and this example clearly illustrates the effect on the beam
size of finite sampling (i.e., only for  $|u|\le u_{\rm max}$).

In practical imaging synthesis, side-lobe levels can be reduced by using
taper function instead of equal weight to all the sampled data.
For instance, by introducing a Gaussian taper function, side-lobe level
can be reduced significantly.
However, while taper is useful for handling side-lobe problems, it is
of no use for improving the main beam size.
In fact, introducing a Gaussian taper function (and other typical tapers
that have lower weight on longer baselines) increases the main beam
size and thus degrades angular resolution of the synthesized image.
This example clearly shows how the standard Fourier transform
approach (with zero-padding) limits the size of synthesized beam and
the angular resolution of image.

\subsection{super-resolution imaging}

In order to obtain super-resolution image, here we consider a different
approach than standard Fourier transform of observed visibilities.
In particular, to avoid zero-padding, we regard observational equations
as underdetermined, which have a series of solutions that are consistent with
observed visibilities.
Figures 1(c) and 1(d) show examples of such a series of solutions.
Here we again assume that the visibility of a point source at $x=0$ is
sampled only for $|u|\le 1$, namely
\begin{equation}
\label{eq:1D-visib-point}
 \mathcal{V}(u)=1\;\; {\rm for}\; |u|\le 1.
\end{equation} 
In figure 1(c) we show several possible visibility functions that
are consistent with observed visibilities in equation
(\ref{eq:1D-visib-point}): $\mathcal{V}=1$ for $|u|<1$, $|u|<2$, $|u|<4$ and 
$|u|<\infty$.
All the four visibility functions are consistent with the observations,
$\mathcal{V} = 1$ for $|u|<1$.
In other words, one cannot distinguish these different functions from
the current observations.
In figure 1(d) we show the Fourier transformation of these four
visibility functions.
As is expected, these different visibility functions result in
${\rm sinc}$-function with different beam-size: when the maximum $u$
becomes larger, the beam becomes sharper.
Here we emphasize that all the four solutions are fully
consistent with observed visibilities (equation
[\ref{eq:1D-visib-point}]), but in the image domain, they provide totally
different angular resolution.
Therefore, limited sample of visibilities such as equation
(\ref{eq:1D-visib-point}) does not necessarily exclude the solutions
with angular resolution finer than the standard beam size.
However, there exist a infinite number of solutions which
satisfy observations, and it is not possible to select a unique solution
solely based on the observed visibilities.
Therefore, to obtain the most-optimized solution, one needs to introduce
additional constraint to select the best one among a series of possible
solutions.

Now the simple question is: how can we select the best solution amongst
infinite number of possible solutions ?
This selection can be done by imposing an additional constraint other than
the observations described in equation (\ref{eq:1D-visib-point}).
In fact, in practical synthesis, additional 
constraint on images are quite often introduced.
For instance, it is often assumed that the intensity distribution
$I(x)$ (or $I(x,y)$ in the two-dimensional case) is non-negative,
except for imaging of absorption features.
Also, in practical radio interferometry synthesis, image is obtained for
a limited area in the sky plane, in which it is implicitly assumed that
the source intensity is distributed within the image size
considered.
In most cases these assumptions are fairly reasonable and often used to
obtain the synthesis images based on the standard Fourier
transformation.

While these assumptions appear somewhat trivial for the standard imaging,
these additional constraints are of great help for selecting the best
solution in the approach considered in the current section.
For instance, in the case for figures 1(c) and 1(d), while 
there are infinite number of ${\rm sinc}$-function solutions of $I(x)$
which are consistent with observations ($\mathcal{V}(u)=1$ for $|u|\le 1$),
an additional constraint on positivity of $I(x)$ automatically select
$I(x)=\delta(x)$ as the best solution.
In this case, the resultant image is a point source which perfectly
matches with the initial image, recovering a super-resolution image with
infinite angular resolution!
Also, if one adopts a constraint on the emitting region,
the best image should be again the point source solution, which has the
lowest side-lobe level and hence provide minimum power outside the area
considered for imaging.
Another possible way to provide an additional constraint is to assume
{``}sparsity{''} of the image.
If one prefers an image with the smallest number of non-zero grids,
again the point source solution becomes the best image amongst infinite
number of ${\rm sinc}$-function solutions.

All the three examples above demonstrate that selecting the best solution with
additional constraint from possible solutions can, at least in some
occasions, provide a super-resolution image which represents the true
image better than that obtained by the standard Fourier transform approach. 
Of course, there is no guarantee if such an approach always works well.
However, one can safely say that while the deterministic approach with
the standard Fourier transformation (with zero-padding) always suffers from
the finite size of the synthesized beam, there is a possibility to obtain
super-resolution images by selecting the best solution in an underdetermined
problem.

\section{Sparse Modeling, Compressive Sensing, and LASSO}

\subsection{Sparse modeling}

As discussed in the previous section, image reconstruction from
interferometry data (visibilities) usually becomes an ill-posed problem.
The zero-padding technique causes a serious degradation of the
image with finite beam size as well as multiple side-lobes.
On the other hand, if zero-padding is not done, basically the
interferometric equation becomes an underdetermined problem for which an
infinite number of solutions exist.
As shown in the last section, a proper selection of the best solution
from possible ones may lead us to super-resolution image.
However, generally speaking the selection is not unique and it is hard
to tell which one would be the best.

Sparse modeling is a class of techniques to solve such an underdetermined
problem effectively.
Basic idea of sparse modeling is to choose the meaningful sparse
solution of underdetermined problem by introducing additional constraint
on sparsity: if most components of
the solution vector are 0 (which is referred to as a {`}sparse
solution{'}), such an underdetermined problem can provide a unique solution.
Since this approach favors sparse solution, one can expect
that its application to interferometer imaging would lead to
super-resolution image, as discussed in the previous section.

\subsection{compressive sensing}

Recently {``}compressive sensing{''} is a very hot topic in the
field of sparse modeling,
and is recognized as a powerful tool to solve such an
underdetermined problem.
Suppose that there is a set of linear equations expressed as
\begin{equation}
 \mathbf{z} = \mathbf{A} \mathbf{x},
\end{equation}
where we assume that the dimension of the observable vector $\mathbf{z}$
(which corresponds to visibility in radio interferometer)
is smaller than the dimension of the solution vector $\mathbf{x}$ (image).
In this case, the above equation becomes an underdetermined problem and 
basically there is no unique solution.
However, if most components of the solution vector $\mathbf{x}$ are 0 (i.e., 
if the solution is sparse), then actual dimension of
observations could exceed the effective dimension of
the solution, and a unique solution will be obtained even though the
apparent dimension of $\mathbf{x}$ is larger than that of $\mathbf{z}$.
This is the basic idea of compressive sensing.

For obtaining such a sparse solution, the most direct way is to minimize
the number of non-zero components of the solution vector.
Generally, $lp$-norm $||\mathbf{x}||_p$ is defined as
\begin{equation}
 ||\mathbf{x}||_p = (\sum |x_i|^p)^{1/p} \;\; {\rm for}\; p>0
\end{equation}
\begin{equation}
 ||\mathbf{x}||_p = ||\mathbf{x}||_0 \;\; {\rm for }\; p=0.
\end{equation}
Note that the $l0$-norm $||\mathbf{x}||_0$ describes the number of non-zero
components of $\mathbf{x}$.
With the use of the $l0$-norm, obtaining a sparse solution from an
underdetermined problem can be regarded as finding a solution with
minimum $l0$-norm, which can be described as
\begin{equation}
\label{eq:sparsemodeling-l0}
 \mathbf{x} = \arg \min ||\mathbf{x}||_0 \;\; {\rm subject\; to}\; 
 \mathbf{z} = \mathbf{A} \mathbf{x}.
\end{equation}
Technically this problem can be solved by trying all possible
combinations of 0 components of $\mathbf{x}$ one by one.
Unfortunately, however, this is computationally very costly,
 and there is no practical algorithm
for solving such a problem for a large dimension.

On the other hand, under some circumstances, it is known that sparse
solution obtained by relaxing $l0$-norm to $l1$-norm 
can provide exact solutions for such a problem, i.e.,
\begin{equation}
\label{eq:sparsemodeling-l1}
{
 \mathbf{x} = \arg \min ||\mathbf{x}||_1 \;\; {\rm subject\; to}\; 
 \mathbf{z} = \mathbf{A} \mathbf{x}.
}
\end{equation}
The condition under which the solution of equation
(\ref{eq:sparsemodeling-l1}) becomes identical to that of equation
(\ref{eq:sparsemodeling-l0})
is revealed by Donoho (2006), and Candes and Tao (2006).
Such a framework for solving underdetermined problems
is quite often referred to as {``}compressive sensing{''} (or sometimes
{``}compressed sensing{''}).
In fact, this has been recognized as a very effective framework to solve
underdetermined problems which could not be solved in a standard manner.
Applications of compressive sensing to radio interferometry are
recently discussed by Wiaux et al.(2009), Wenger et al.(2010), Li et
al.(2011), and so on.
Compressive sensing and sparse modeling will be used in a wider field of
astronomy (e.g., \citet{Starck10}), as seen in its rapid growth in
other fields of imaging science such as MRI (\citet{Lustig08}).

\subsection{LASSO}

LASSO (Least Absolute Shrinkage and Selection Operator ) is another
widely-used technique in sparse modeling, originally developed in the
field of statistics (Tibshirani 1996). 
As is the case for compressive sensing, LASSO also utilizes $l1$ norm,
but the solution $\mathbf{x}$ is obtained as,
\begin{equation}
\label{eq:lasso}
 \mathbf{x} = \arg \min \left( ||\mathbf{z}-\mathbf{A} \mathbf{x}||_2^2 +
 \Lambda ||\mathbf{x}||_1 \right),
\end{equation}
where $\Lambda$ is a regularization parameter.
The first term in the argument of the right-hand side describes goodness
of fit, which is 
commonly used in least-squares fitting, and the second term is the
penalty term based on the $l1$-norm.
The regularization parameter $\Lambda$ adjusts degree of sparsity by
changing the weight of the $l1$-norm penalty.

The underlying idea of compressive sensing and LASSO is quite similar:
for any underdetermined problem, there are infinite number of solutions which satisfy the conditions such as $\mathbf{z}=\mathbf{A}\mathbf{x}$ or $||\mathbf{z}-\mathbf{A}\mathbf{x}||_2^2< \epsilon$.
The key issue is how to select the most realistic solution (which best represents the truth), and both compressive sensing and LASSO do this by selecting sparse solution which has many zero components in $\mathbf{x}$.
The difference between the two is that in compressive sensing the selection is simply done only by $l1$-norm and thus selection is rather automatic, but in LASSO one can adjust the sparsity of the solution by varying the regularization parameter $\Lambda$.
For instance, a large $\Lambda$ prefers a solution with very few non-zero
components, while $\Lambda=0$ introduces no sparsity (in which case one cannot choose any specific solution from infinite number of the possible solutions).

Generally optimization of $\Lambda$ in LASSO is an important issue. 
Some statistical methods, such as information criteria or cross
validation, can be used for this optimization. 
Another idea is to adjust $\Lambda$ based on the
estimate of the sparsity, which might be obtained from some a
priori information or theoretical predictions.
Optimized value of $\Lambda$ is dependent of characteristics of
problems to be solved (i.e., the shapes of observation matrix $\mathbf{A}$
and also image $\mathbf{x}$), and the detailed discussion on this issue
is to be given in a future work.
In this paper, we basically use $\Lambda$ which is an order of unity, but in section 5 we also discuss the effect of $\Lambda$ in the resultant images of black hole shadow.

We also point out that when applying compressive sensing or LASSO, it is important that the phenomena which we are dealing with indeed have sparse solution. Otherwise, introduction of sparsity becomes unrealistic, and thus the sparse solution obtained by compressive sensing or LASSO is of no use.
However, as we discussed later, in the case of VLBI imaging, where only brightest part of the emission is observable due to the filtering effect of interferometer (i.e., extended emissions are resolved out), we can expect sparsity in VLBI images, and thus one can expect that such an approach could work, in particular for imaging black hole shadow.

\subsection{numerical simulation}

In this paper we formulate the interferometric imaging problem with
LASSO as follows, 
\begin{equation}
  \label{eq:lasso2}
  \mathbf{x} = \mbox{argmin} (\|
  \mathbf{z}-A\mathbf{x}\|_2^2+\Lambda\|\mathbf{x}\|_1),
  ~~\mbox{subject to} ~~\mathbf{x}\ge \mathbf{0},
\end{equation}
where the inequality applies componentwise. 
Here we add a positivity condition for $\mathbf{x}$ because
our major interest is to image emission rather than absorption.
Note that in the current paper $\mathbf{x}$ in equation
(\ref{eq:lasso2}) corresponds to an image vector.
For two dimensional cases, its dimension is 
$N_{\rm grid}^2$, where $N_{\rm grid}$ is the number
of grids along $X$ and $Y$ axes of images.
Also, $\mathbf{z}$ corresponds to the observed visibility vector
$\mathcal{V}(u,v)$, and the observation matrix $\mathbf{A}$ consists of the
exponential components in the Fourier transformation, i.e., 
$A_{u,v}=e^{-2\pi i(ux + vy)}$.

In sparse modeling, an important question is in which domain the
solution becomes sparse.
If the real solution is not sparse in the domain under consideration,
applying sparse modeling (compressive sensing or LASSO) could lead to
unreasonable solutions which do not represent the real solution.
In the present study, we simply assume that the solution
becomes sparse in the image domain, because we focus on
observations of super-massive black holes with mm and sub-mm VLBI.
We note that the emissions observed with mm and sub-mm VLBI are
compact because VLBI arrays intrinsically work as a {`}filter{'} to
extract compact high-brightness components (extended emissions are
resolved out).
Such compact emissions are expected to be mostly
associated to the vicinity of black holes and thus confined to a part of
an image area, making the expected image sparse in the image domain itself.

We solve the problem described in equation (\ref{eq:lasso2})
by rewriting this as a
Quadratic Programming (QP) problem and use QP solver. 
The standard form of QP is 
\begin{equation}
  \label{eq:QP}
  \mbox{min}
  \frac{1}{2}\mathbf{x}^TQ\mathbf{x}+\mathbf{c}^T\mathbf{x},
  ~~~\mbox{subject to}~~ D\mathbf{x}\le \mathbf{b}, 
\end{equation}
where $^T$ denotes the transpose, $\mathbf{c}$ is a column vector, and $Q$ is
a symmetric matrix. 
It is easy to rewrite equation (\ref{eq:lasso2}) as equation
(\ref{eq:QP}). 
In our numerical simulations, we used MATLAB QP solver to
solve the one-dimensional problem. 
For one-dimensional problem, we used a naive
iterative-shrinkage-thresholding (IST) algorithm (Daubechies, Defrise \&
De Mol 2004). 
Note that essentially both methods give identical results.

\begin{figure}[t]
\begin{center}
       \FigureFile(80mm,80mm){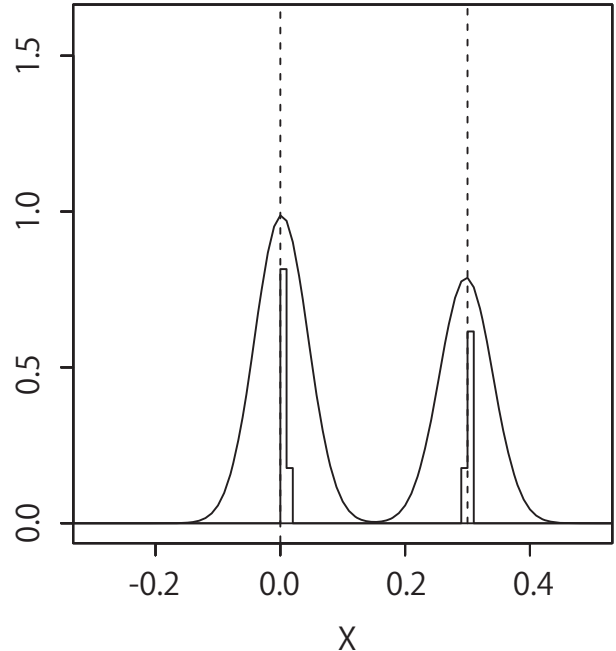}
\end{center}
\vspace{-0.5cm}
\caption{A simulated one-dimensional image of two adjacent point sources
 solved by LASSO (for $\delta x$=0.3). 
 Histogram corresponds to the solution by LASSO, and solid
 curve corresponds to the one-dimensional image convolved with one-tenth
 of the standard synthesize beam (i.e., a full width of $\sim 0.1$).
 Vertical dashed lines denote the true
 positions of the two point sources.}
\label{fig:two-points}
\end{figure}

\section{One-dimensional simulations}

\subsection{two point sources}

In order to see how super-resolution image can be achieved by the
technique of sparse modeling, first we consider the simplest
example, namely a problem to identify two adjacent point sources in 
one-dimensional space.
We note that in such a case visibility $\mathcal{V}(u)$ and image $I(x)$ are
related to each other through equation (\ref{eq:1D-FT}).
Here we assume two point sources within the standard synthesized beam, 
with the main component at the origin, and the
secondary component at $x=\delta x$.
We set intensity of the main component to be unity and that of the
secondary to be 0.8 (the intensity difference is introduced to test if the
amplitudes of the two components are retrieved well).
In mathematical expression, we may write as
\begin{equation}
 I(0) = 1, \;\;
 I(\delta x) = 0.8, \;\;
 I = 0 {\rm\;\; for\; other\;\;}x
\end{equation}
We also assume one dimensional observation which uniformly samples the
$u$-plane for $|u|<1$.
This setting corresponds to a synthesized beam size of unity in
the image domain, and we try to resolve two point sources with $\delta
x<1$ using sparse modeling.
Here the solution $I(x)$ is obtained from the sampled visibilities $\mathcal{V}(u)$
by using the LASSO algorithm shown in equation (\ref{eq:lasso2}).

Figure 2 shows an example of the simulation results for $\delta x=0.3$.
As seen in figure 2, the two point sources within the standard beam size
(which has a full width of $\sim 1$ in $x$) are clearly separated in the
image obtained with LASSO.
This simple simulation clearly illustrates that structure within the
beam size can be resolved by the sparse modeling technique (but
note that noise is absent in this simulation. 
The effect of the noise will be discussed later in section 4.3).
The dashed lines in figures 2 denote the initial positions of the main and
secondary point sources, showing that the position of the secondary
peak is accurately retrieved.
Also, the intensity radio of the two sources is reproduced well.
The results shown here clearly demonstrate that obtaining images with
sparse modeling can indeed retrieve information on structure smaller
than the synthesized beam size, as is discussed in the previous section.
For a larger value of $\delta x$, the sparse modeling also works
well just as the case shown in figure 2.
On the other hand, for $\delta x$ smaller than $\sim 0.1$, the solution
becomes a single component located between the two point sources in the
initial image,
indicating that there is a critical separation below which separating
the two adjacent points becomes difficult.

\subsection{optimal restoring beam-size}

\begin{figure}[t]
\begin{center}
       \FigureFile(85mm,85mm){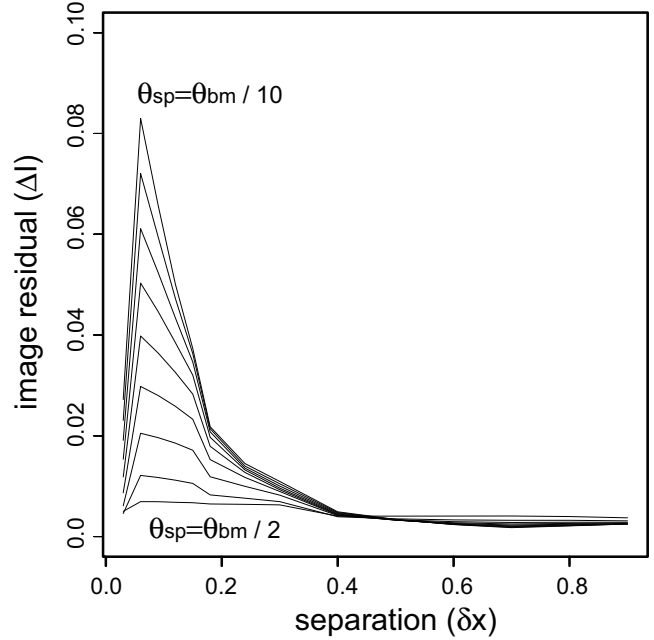}
\end{center}
\caption{Image residuals $\Delta I$ as a function of the two point separation
 $\delta x$ with
 varying the restoring beam-size of super-resolution image.
From the bottom to top, the restoring beam size is 1/2, 1/3, ...,1/10 of
 the synthesized beam size (i.e., super-resolution factor of 2, 3, ..., 10.).}
\end{figure}

As we have seen above, sparse modeling can provide information in a scale
finer than the synthesized beam size.
On the other hand, there is a separation limit in resolving the two
points, and hence super-resolution image should also have a finite beam size.
Then, here comes the next question: what is the optimal size of the restoring
beam for such super-resolution images ?
To evaluate this, here we used the simulation in the previous section, and
compared the true (initial) image and reconstructed image with varying the
restoring beam size.
Suppose that the initial image is $I_0(x)$ and the solution by
sparse-modeling is $I_{\rm sol}(x)$.
By introducing a Gaussian restoring beam function 
$g(x)\propto \exp(-x^2/2 \sigma)$ where 
$\sigma=\theta_{\rm sp}/2\sqrt{2\ln 2}$,
the restoring-beam-convolved initial image is given by
\begin{equation}
 I_{\rm init}(x)= \int g(x-x',\theta_{\rm sp})\times I_0(x') dx',
\end{equation}
and the super-resolution image is given by
\begin{equation}
 I_{\rm obs}(x)=\int g(x-x',\theta_{\rm sp})\times I_{\rm sol}(x') dx'.
\end{equation}
Here $\theta_{\rm sp}$ is FWHM of the restoring beam size for
super-resolution image.
With these, one can define image residual, which measures the goodness of the
resultant super-resolution image, as
\begin{equation}
 \Delta I =\frac{\int |I_{\rm init}(x)- I_{\rm obs}(x)| dx}{\int I_{\rm
  init}(x) dx}.
\end{equation}

Figure 3 shows $\Delta I$ as a function of the separation of the two
point sources in the initial image ($\delta x$), with nine different
values of restoring beam size, $\theta_{\rm sp}$ =$\theta_{\rm bm}/2$,
$\theta_{\rm bm}/3$, ...,  $\theta_{\rm bm}/10$ ($\theta_{\rm bm}$ is the
standard beam size, which is equal to unity here).
As seen in figure 3, the largest difference between the initial image
and super-resolution occurs around $\delta x=0.05$, where the identification
of the two components becomes difficult.
However, even for the worst case with $\delta x=0.05$, the image
residual $\Delta I$ remains within 10\% for a restoring beam size of
$\theta_{\rm bm}/10$ (or super-resolution factor of 10).
If one requires the maximum image residual at $\sim$3\% level or less, 
then the super-resolution factor of 5 is still acceptable.
Therefore, these simulation results indicate that based on the sparse
modeling technique super-resolution image by a factor of a few or slightly 
more is achievable within reasonably small errors.

\subsection{Effect of noise}

\begin{figure*}[t]
\vspace{-3cm}
\begin{center}
       \FigureFile(170mm,100mm){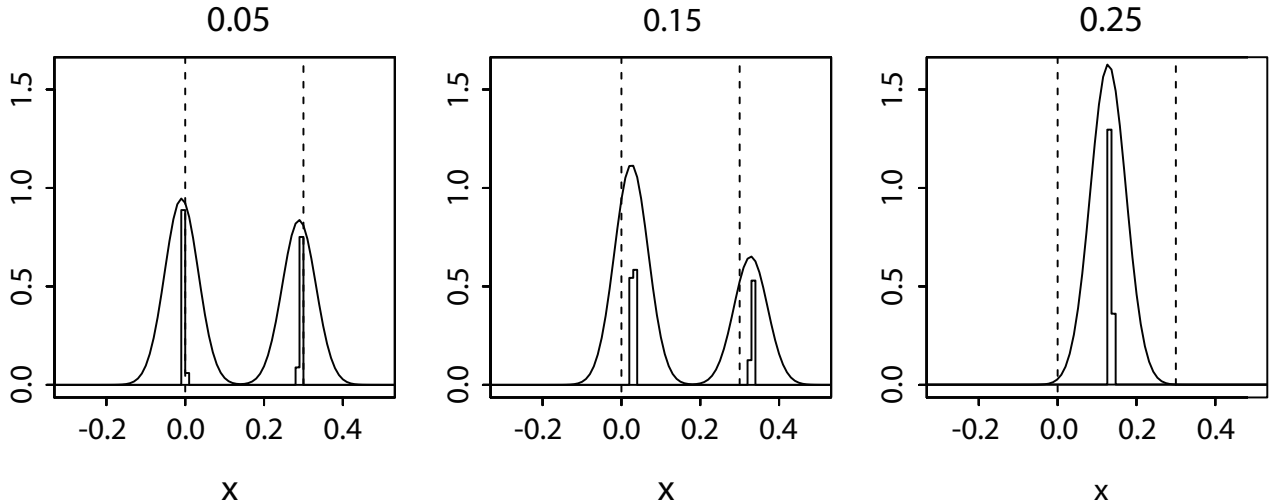}
\end{center}
\vspace{-2.5cm}
\caption{
LASSO solutions for the two point-source image with various noise level.
Here the simulation condition is the same to that in figure \ref{fig:two-points}, with the separation of the secondary point at $x=0.3$ and the intensity ration of 0.8. Here Gaussian noises are added to the observed visibilities, with a Gaussian noise $\sigma$ equivalent to 0.05, 0.15 and 0.25 of the peak visibility amplitude (indicated at the top of panels). For the small noise case ($0.05$ of the peak), the reconstruction with LASSO is nearly perfect. In the modest noise case, the two points are still separated but the positions and intensity ratio are moderately affected. For the largest noise of 0.25, the two points are merged into one component and cannot be resolved due to the effect of noise.}
\label{fig:one-dim-with-noise}
\end{figure*}

Here we consider the effect of noise, which was not considered in the above discussion.
We solve the same problem shown in figure \ref{fig:two-points} with the source separation of $\delta x=0.3$, but this time we added Gaussian noises to the visibilities, and the image reconstruction is done in the same manner using LASSO.
Figures \ref{fig:one-dim-with-noise} show the results for the three cases with different noise amplitude, with Gaussian noise level $\sigma$ to be 5\%, 15\%, and 25\% of the peak visibility amplitude.
For the Gaussian convolution, again we used the one-tenth of the standard beam size.

For the case with the 5\% noise level, the image reconstruction is nearly perfect: basically the two peaks are recovered correctly in terms of both the positions ($x=0$ and $0.3$) and the peak flux (1 and 0.8).
This result ensures that as far as the noise is sufficiently small, the addition of the noise does not alter that the main conclusion that the image reconstruction with LASSO can retrieve the structure finer than the beam size.
On the other hand, the middle panel of figures \ref{fig:one-dim-with-noise} corresponds to the case with a noise level of 15\%.
In this plot, one can see that larger noise gradually affects the result, as the flux ratio of the two peaks become different from the initial values (originally 1 to 0.8), but more power is concentrated to the component at $x=0$.
Also the positions of the two peaks are slightly shifted by $\sim 0.03$, but this is still fairly smaller than the beam size itself.
Hence such a super-resolution image may be still acceptable depending on what kind of conclusion is drawn from the image.

In the right panel of figures \ref{fig:one-dim-with-noise}, we show the case with 25\% noise level (which corresponds to a case in which SNR in only 4 even at the visibility peak).
In this case, the effect of noise is significant, as the two peaks are merged into one, and there is no sign of the double peak structure.
Above simulation results show: 1) that the super-resolution technique proposed in this paper can be of use for practical cases in which the observational noise exists, and 2) that the super-resolution image requires relatively good SNR ratio, as easily expected.

For other separations, basically smaller separation case becomes more sensitive to noise: for instance, in case of $\delta x=0.15$, addition of noise larger than 3\% level would make it impossible to separate the two peaks.
Therefore, very high super-resolution factor is not practical in terms of the tolerance against noise.
However, as we have seen in figure \ref{fig:one-dim-with-noise}, super-resolution factor of a few is applicable even under the existence of noise.

It is remarkable to point out that the solutions by LASSO shown in figure \ref{fig:one-dim-with-noise} have a baseline which exactly matches zero other than at the location of the signal (around $x=0$ and $0.3$), without any noise floor.
This apparent cancellation of the noise is also due to the sparse modeling:
while solutions with noise floors could have moderately smaller value in the error term $||\mathbf{z}-\mathbf{A}\mathbf{x}||_2^2$ (the first term in the argument of the LASSO equation), such solutions with noise floor should have a large number of non-zero components in $\mathbf{x}$, and thus they are discarded due to the large penalty term, $\Lambda||x||_1$.
As such, the preferred solution by sparse modeling tends to have zero baseline and thus noise is apparently reduced (though it is not necessarily clear if the apparently noiseless image is indeed a better representation of the reality or not. Since this is beyond the scope of our paper, we leave this issue for future studies).

\section{Two-dimensional simulations}

\begin{figure}[t]
\begin{center}
       \FigureFile(85mm,85mm){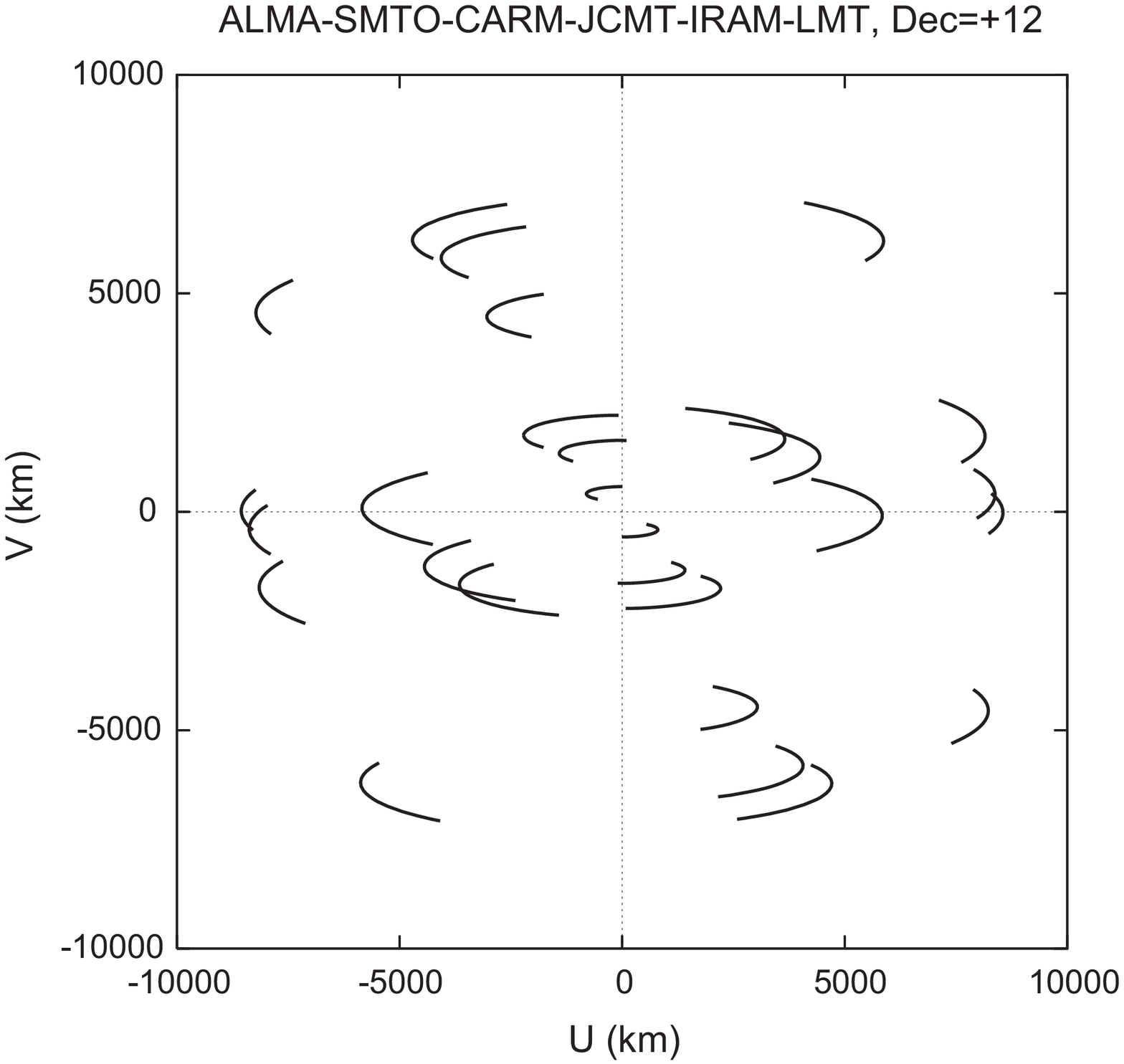}
\end{center}
\caption{Simulated UV coverage of M87 with six-station sub-mm VLBI
 array of EHT. Here it is assumed that observations are conducted at an
 elevation larger than 20$^\circ$ at each station.}
\label{fig:UV}
\end{figure}

\begin{figure*}[t]
\begin{center}
       \FigureFile(160mm,100mm){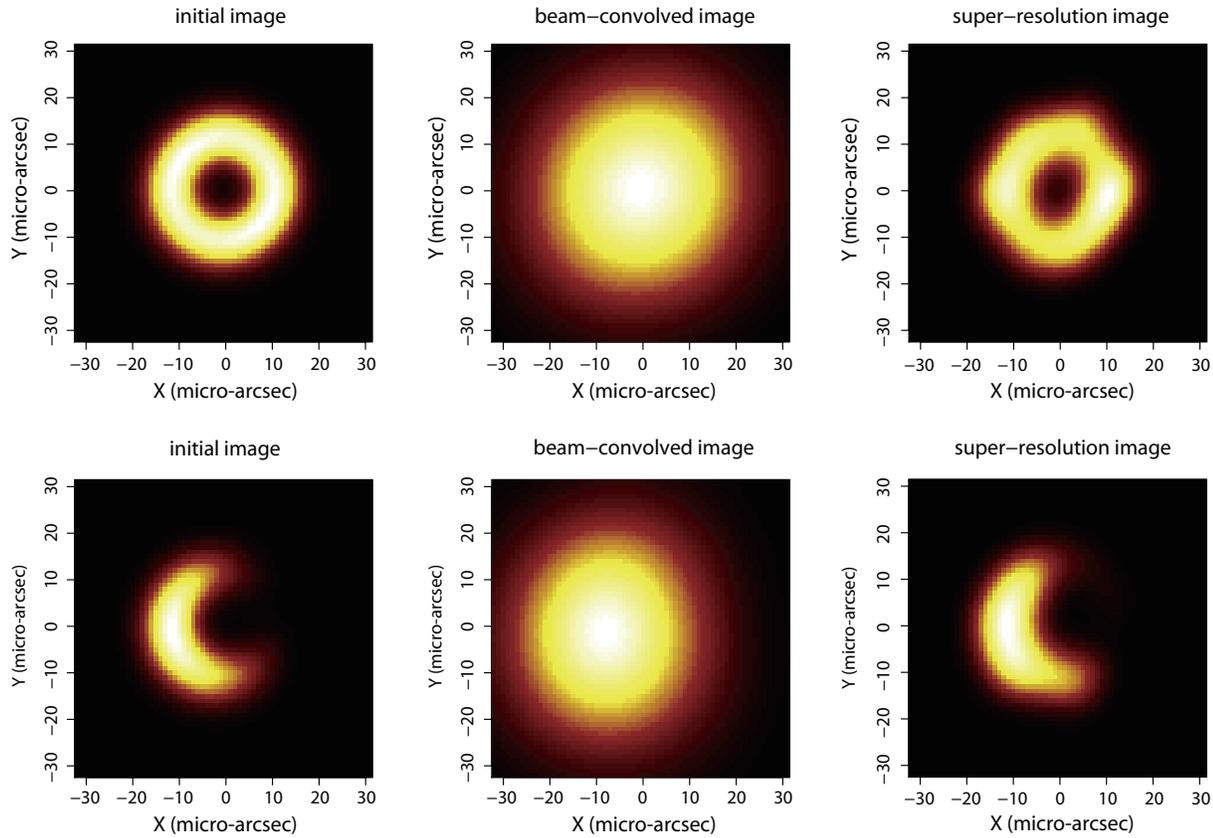}
\end{center}
\caption{Imaging results for simulated EHT observations of M87's black hole
 shadow. The upper panels correspond to the ring case, and the lower
 panels to the crescent case. From the left to the right, the panels
 show an initial
 image, convolution with the standard synthesized beam, and the solution
 with sparse modeling.
 Each image has 64$\times$64 grids with a grid size of 1$\mu$as.
 The initial images as well as super-resolution reconstructed images are
 convolved with a restoring beam that is finer by a factor of 4 than the
 standard synthesized beam.}
\label{fig:BH-shadow}
\end{figure*}

Now we discuss two dimensional case, which corresponds to actual
imaging of astronomical objects using radio interferometer.
In particular, here we consider an application of the super-resolution 
technique to VLBI observations of black hole shadow, 
in which obtaining high resolution is fundamental.

\subsection{simulation conditions}

Direct imaging of black hole shadow will provide the ultimate
confirmation of existence of super-massive black holes at the centers
of galaxies.
To build an array which enables us to resolve black hole shadows,
now the EHT collaboration is developing a world-wide sub-mm VLBI
array.
Here we simulate EHT observations that are expected to become
real in near future.
Currently three sites in US (CARMA: Combined Array for Research in
Millimeter-wave, SMTO: Submillimeter Telescope Observatory, JCMT: James
Clerk Maxwell Telescope/SMA: Submillimeter Array/CSO: Caltech
Submillimeter Observatory) are operational at
1.3 mm and producing scientific results on regular basis (e.g., Doeleman et
 al. 2008; 2012; Lu et al. 2012; 2013).
In the present paper, we consider an array with six stations including
additional three stations which are to be realized in near future,
namely, ALMA (Atacama Large Millimeter/submillimeter Array) in Chile, 
LMT (Large Millimeter Telescope) in Mexico, and IRAM (Institute de
Radio Astronomie Millim\'{e}trique) 30m in Spain.
All the latter three stations have concrete plans for participating
sub-mm VLBI observations (ALMA and LMT) or even already have test observations
(IRAM).
Thus the six-station array considered here is likely to be
operational soon.
As for the observing wavelength, we take $\lambda=1.3$ mm (or a frequency
$\nu$ of 230 GHz), which is more easily available than a shorter
wavelength such as $0.8$ mm, although both wavelengths are within the 
scope of the EHT project.

For the target source, here we consider M87, which is one of the best
target along with the Galactic center black hole Sgr A* in terms of 
apparent shadow size. 
The mass of M87's black hole is still matter of debate: while
Gebhardt \& Thomas (2009) suggested a mass of 6.4($\pm$0.5) 
$\times 10^9 M_\odot$, recent re-evaluation of the black hole mass
gives a mass of
$3.5\times 10^9 M_\odot$ (Walsh et al. 2013, see also Macchetto et
al. 1997 for earlier studies suggesting a small mass).
The latter is smaller by a factor of two than the previous estimate by
Gebhardt \& Thomas (2009).
If one adopts a mass of $3.5\times 10^9 M_\odot$ and a distance of 17
Mpc, the expected diameter of the black hole shadow is around 20 $\mu$as
for a non-rotating black hole, and it could be smaller for a spinning
black hole.

Figure \ref{fig:UV} shows EHT's UV-coverage plot for M87 (at a source
declination of $\delta=+12^\circ$).
Here we assume that the source is observed for an elevation angle beyond
20 deg at each station.
As seen in figure \ref{fig:UV}, the maximum baseline length is around
9000 km, and hence for an observing wavelength of 1.3 mm, this array
provides an angular resolution of $\sim 30$ $\mu$as.
In this case, the angular resolution (the standard beam size) could
be still larger 
than the expected size of the shadow diameter (in particular for the
case of small mass and/or high spin black hole).
Hence it could be
difficult to resolve the shadow based on the standard imaging synthesis in
case of the small mass.
Therefore, it is of great interest to test if the super-resolution technique 
effectively boost the image resolution to resolve the black hole shadow.

\subsection{results}

We conducted simulations for two cases
of the black hole shadow of M87: a ring-like shadow and a crescent
shadow (both with a diameter of 20 $\mu$as).
Here, in order to emulate real observations, we assume that visibilities
are sampled every 10 minutes with the ($u,v$) coverage shown in figure
\ref{fig:UV}, and also random noises are added to
visibilities at a 5\%-level of the peak flux: in order words,
visibility SNR is 20 for the shortest baseline and smaller for longer
baselines (visibility amplitude decreases with baseline length as source
structure is partially resolved out with longer baseline).
The images are reconstructed by applying LASSO to the simulated
visibilities (as described in section 3.4), and convolved with a
super-resolution beam which is smaller than the standard synthesized
beam by a factor of 4.
Note that in this simulation there are 976 visibility samples.
Since the solution dimension (the number of image grids) is $64^2=4096$,
the problem studied here is in fact an underdetermined one.

In figures \ref{fig:BH-shadow}, we show the results for the ring-like
shadow in the top panels, and for the crescent shadow in the bottom panels,
respectively.
All the images have 64$\times$64 grids, with a grid size of 1 $\mu$as. 
The left panels of figures \ref{fig:BH-shadow} show initial model images
that are convolved with a super-resolution beam which is four times
finer than the standard synthesized beam size.
The middle panels are images convolved with the standard synthesized beam
(the synthesize beam size is 33$\times$29 $\mu$as with a position angle
of 30 deg).
The right panels show images reconstructed with LASSO.

The right panels of figures \ref{fig:BH-shadow} clearly show the
super-resolution images using sparse modeling indeed reproduce the
detailed structure within the synthesized beam: while the shadow
structures cannot be seen in the standard-beam-convolved images (in the middle
panel of figures \ref{fig:BH-shadow}), in the super-resolution images
the ring/crescent structures are clearly reconstructed.
Of course, the existence of the noise in the present simulation avoids
the perfect matches between the initial image and reconstructed image.
However, basic structures of brightness distribution are traced well.

\begin{figure}[t]
\begin{center}
       \FigureFile(80mm,80mm){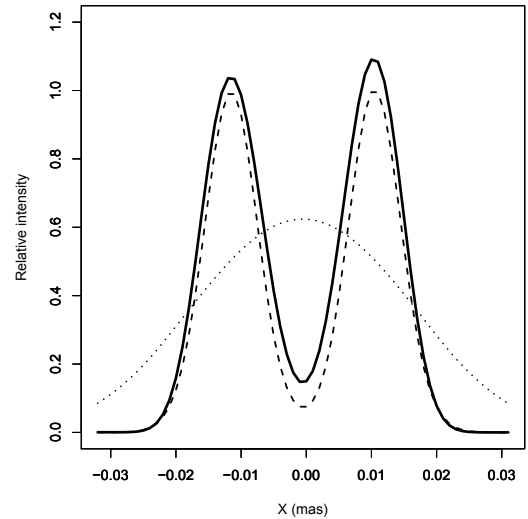}
\end{center}
\vspace{-0.5cm}
\caption{Cross-section of the ring-like shadow image in
 figure \ref{fig:BH-shadow}) (at $Y=0$). 
 Thick curve represents the super-resolution reconstructed image, and dashed
 curve is the initial image, both of which are convolved with the
 super-resolution beam, which is 4 times finer than the standard synthesized
 beam. Dotted curve shows the image convolved with the standard
 synthesized beam, where the double peak structure of the ring is
 totally lost due to the convolution effect.}
\label{fig:shadow-cross-section}
\end{figure}

To clarify the advantage of the super-resolution technique over the
standard images, in figure \ref{fig:shadow-cross-section} we show the cross section of the
ring results (the top panels of figures \ref{fig:BH-shadow}) along $Y=0$.
As seen in figure \ref{fig:shadow-cross-section}, the twin peaks (corresponding to the ring) are well-reproduced in the super-resolution image.
On the other hand, in the standard imaging, the two peaks are merged into a
single component due to the convolution effect, and it is difficult to
retrieve the two peaks from the convolved image based on the standard
CLEAN algorithm.
Here we note that in the 
CLEAN process, CLEAN component is always set at the peak in the
convolved image, and hence the CLEANed imaged should have a single peak at 
$X \sim 0$ rather than two peaks at $X \pm \sim 10$ $\mu$as.
Therefore, in the cases shown in figures \ref{fig:BH-shadow}, the
super-resolution technique using sparse modeling indeed has strong
advantage against the standard imaging, and
the results here demonstrate that the super-resolution technique will be of
great help to resolve the shadow of super-massive black holes.

\begin{figure*}[t]
\begin{center}
       \FigureFile(170mm,80mm){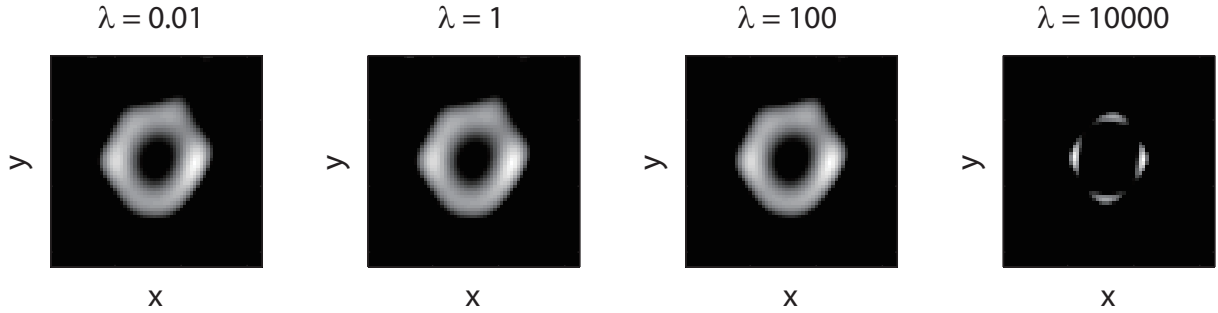}
\end{center}
\vspace{-0.5cm}
\caption{Comparison of two-dimensional super-resolution images reconstructed by LASSO, with different values of $\Lambda$. The conditions are identical with those shown in the upper panel of figures \ref{fig:BH-shadow}, except that here the convolution with the super-resolution beam is not conducted so that the direct comparison between the solutions could be easier.
 From left to right, $\Lambda$=0.01, 1, 100, and 10000. The left three figures provide nearly identical results, while the rightmost one shows an image too much simplified due to very strong constraint of the sparsity.}
\label{fig:BH-shadow-with-lambda}
\end{figure*}

\subsection{Choice of $\Lambda$}

In order to evaluate how the resultant image varies with a choice of $\Lambda$, here we solve the same problem in the previous section by varying $\Lambda$.
Figures \ref{fig:BH-shadow-with-lambda} show the ring cases (corresponding to those in the upper panels of figures \ref{fig:BH-shadow}) with four values of $\Lambda$, 0.01, 1, 100, and 10000.
Note that the image area has the same size with those in figures \ref{fig:BH-shadow}.
Also note that in figures \ref{fig:BH-shadow-with-lambda}, the convolution with the super-resolution beam size, which is included in figure \ref{fig:BH-shadow}, is not taken into account, so that the difference between the solutions can be directly compared in detail.

For $\Lambda$ smaller than 100, basic trend is common: the ring-like structure within the standard beam size is recovered well.
On the other hand, for very large $\Lambda$ case ($\Lambda=10000$), the $l1$ regularization term imposes very strong sparsity in the solution, and hence most of the power is forced to concentrate to four arclets along the ring-like structure.
Certainly this result shows that introduction of too much sparsity could result in non-realistic image which is over-simplified.
However, for modest range of $\Lambda$, the LASSO solution provides reasonable representation of the initial model.
If a priori information on the source is available (such as rough idea on how the source image appear), then such information would be helpful for discriminating the two different types of solutions in figures \ref{fig:BH-shadow-with-lambda}.
Or, even if there is no a priori information on the possible source structure, the LASSO solutions with wide range of $\Lambda$ can still provide a few possible classes of solutions (such as the one for $\Lambda=0.01/1/100$ and the one for $\Lambda=10000$), which is still very helpful for constraining the source structure.

\section{Relevant future issues}

Here we briefly discuss issues related to this study, especially those
important when implementing the super-resolution technique to real data
analysis of mm and sub-mm VLBI.

\subsection{Sources with extended structure}

In the present study we focused on black hole shadow images in which
emissions are confined to the vicinity of the black hole.
Such an image is expected to be sparse in the image domain, and hence
the sparse modeling technique works successfully
to reconstruct images with super-resolution.
The key here is that mm/sub-mm VLBI array, such as EHT, has relatively
low sensitivity and sparse $uv$ coverage, and hence the observation works
as a 'filter' to extract compact and bright components from real images
(in other words, extended structures are resolved out).
In reality, however, black holes could have extended structure such as
accretion disk and jets.
With technical development in VLBI as well as sensitivity
upgrade of the array, relatively faint and extended structure would
become detectable in near future.
Then, the assumption of {``}sparsity in the image domain{''} would no
longer hold, and the sparse modeling should be modified in order to
reconstruct the true image.

How can we treat such a none-sparse image with the technique of 
sparse modeling?
As far as black hole shadows are concerned, one way to handle such a
problem is to separate extended components and the compact component
associated with the black hole, to the latter of which one
would like to apply the super-resolution technique.
This approach can be realized with following steps :
first, apply the standard imaging (Fourier transformation with zero-padding
and CLEAN) to the data and obtain the map
including extended structure and the compact component (which are 
convolved with the standard synthesized beam).
Secondly, the contribution of extended structures are subtracted from
the observed visibilities.
This can be done by subtracting visibility contribution of the CLEAN
components of extended structure.
The remaining visibilities are associated only with the compact component and
thus now the image is expected to be sparse (without extended
structures). 
At this stage one can apply the LASSO to obtain a super-resolution
image of the black hole shadow, as we have already demonstrated in
the previous sections.

Another possible approach is to modify the regularization term in
equation (\ref{eq:lasso}).
For instance, in the field of imaging, the regularization term based on
 total variation (TV: $\propto \sum \left|x_i - x_{i+1}\right|$) is often
 used.
The regularization with TV prefer smoother distribution of power with
less discreteness.
Thus, inclusion of the TV term could lead to better imaging of extended
structure, but a trade-off is that this would lower resolution
compared to that of the regularization only with $l1$-norm.
To optimize the image, combination of the two regularization term 
could be used for tracing both compact and extended sources.

In more general cases, one might be interested in applying the sparse
modeling technique to astronomical objects other than black holes.
In such a case, the main reasoning of applying the sparse modeling is
not super-resolution, but there is still significant merit because
solving an underdetermined problem without zero-padding avoids generation
of side-lobes, and it avoids process like
CLEAN which often requires non-automated interactive analysis.
Previous works by \citet{Wiaux09} and \citet{Li11} discussed such a
general application.
We note that in such a case, there is no guarantee
that the image is sparse in the image domain, and hence the LASSO and
compressive sensing may not work in the manner presented here.

To overcome this, it is fundamental to transform the image
to another domain
in which the corresponding {`}solution{'} becomes sparse. 
One of such possible transformation is wavelet, as the wavelet
transformation is known to make natural images sparse in the wavelet
domain (and hence it is used for image compression).
In fact, \citet{Li11} intensively studied imaging by isotropic decimated
wavelet transform, and demonstrated its advantage over the standard
CLEAN process.
However, we note that if the sparse solution is obtained in a different
domain such as wavelet, it is not guaranteed if the resolution
could be better than the standard imaging, and hence it may not be an
effective approach for black hole imaging.

\subsection{case for non-linear problem}

The discussion in this paper so far implicitly assumed that we can
obtain both visibility amplitude and phase, i.e., complete visibilities
as complex.
In normal interferometer such as connected arrays and cm VLBI arrays,
this is often the case.
However, for sub-mm VLBI there is possibility that low sensitivity
and/or poor weather condition prevent from obtaining complete visibilities
(usually phase is more difficult to obtain).
In such a case, an alternative quantity (less complete but far better than
nothing) is
bi-spectrum, which is a triple product of visibilities from three triangular
baselines, i.e.,
\begin{equation}
\mathcal{V}_{123}=|\mathcal{V}_{12}||\mathcal{V}_{23}||\mathcal{V}_{31}| e^{i(\phi_{12}+\phi_{23}+\phi_{31})}. 
\end{equation}
Here subscripts 1, 2 and 3 denote stations in the triangle, and 
set of two subscripts denotes the relevant baseline with the two stations.

Bi-spectrum (or closure phase when its phase is concerned) is relatively
easy to obtain compared to individual phases such as $\phi_{12}$.
In fact, even when visibility phase is difficult to obtain, in VLBI
observations, bi-spectrum can be measured based on incoherent averaging
(e.g. \citet{Rogers95}).
Similarly, bi-spectrum are available with optical/infrared interferometer,
and its use in image reconstruction has been investigated (e.g., BSMEM,
an image reconstruction from bi-spectrum with maximum entropy method,
see \citet{Buscher94}).

If the only available phase is the closure quantity, one cannot write the
observational equations (the two-dimensional Fourier transform) in
linear equations, and thus the sparse modeling such as
compressive sensing/LASSO cannot be applied.
This would severely limit the application of the technique discussed in
the present paper.

To overcome this limitation, an iterative approach could be used.
As a first step, one creates an initial model image based on the
observed bi-spectrum and/or a priori knowledge of the astronomical object.
Then, by combining the initial image with observed bi-spectrum,
visibilities (amplitude + phase) are estimated for each baseline.
Once full set of visibilities are estimated, one can apply sparse
modeling to
the estimated visibilities to obtain an image.
Of course, the obtained image may not be a good representation of the
true image because the visibilities are estimated from bi-spectrum and
the initial model image.
To improve the image, one estimate the complete visibilities again based
on the new image and observed bi-spectrum, and apply the sparse modeling
again to refine the image.
Such procedures are conducted iteratively until the image converges.

Such an iterative technique is used in the analysis of X-ray
diffraction images, in which the basic equation is the same to the
radio interferometer, but phases are never observed (the technique is
called as {``}phase retrieval{''}, e.g., \citet{Ikeda12}).
Therefore, the iterative treatment would be helpful to find a good
solution which represents the observed bi-spectrum well, as long as the
initial image is properly chosen.
Testing whether or not this scheme works for black hole imaging would be
of great interest and will be a good target for near future work.

\section{Summary}

In the present paper, we presented super-resolution imaging technique
based on sparse modeling.
Usually interferometric imaging is underdetermined problem that
has an infinite number of possible solutions.
We showed that selecting one solution from many possible solutions can
provide better resolution than the standard images obtained with
{``}zero-padding{''}, 
and we demonstrated that sparse modeling, in particular LASSO in this paper,
can be used to select a solution based on the sparsity of the solution.
Based on the simulations, we found that the above scheme works well both for
one-dimensional and two-dimensional cases, and especially demonstrated
its power when applied to VLBI observations of black hole shadows.
Therefore, the super-resolution technique will hopefully have a strong
impact in future observations with radio interferometry.

\bigskip
The authors would like to thank for Masato Okada, Kenji Nagata, 
Hiroshi Nagai and Daisuke Kuroda for fruitful discussions.
We are also grateful to the referee, Prof. J. Moran, for his careful reading and constructive suggestions which improved the paper.
The authors also acknowledge financial support by the MEXT/JSPS KAKENHI
Grant Numbers 24540242, 25120007 and 25120008.

\end{document}